\documentclass[sigconf,nonacm]{acmart}
\usepackage[backend=bibtex,style=verbose-trad2,style=numeric,
citestyle=numeric]{biblatex}
\usepackage{amsmath,amsfonts}
\usepackage{subcaption}
\usepackage{algorithmic}
\usepackage{graphicx}
\usepackage{textcomp}
\usepackage{xcolor}
\usepackage{listings}
\def\BibTeX{{\rm B\kern-.05em{\sc i\kern-.025em b}\kern-.08em
    T\kern-.1667em\lower.7ex\hbox{E}\kern-.125emX}}
\usepackage{sansmath}
\usepackage{hyperref}
\usepackage{tikz}
\usepackage{float}
\usepackage{balance}
\usepackage{soul}

\bibliography{papers}
\newcommand{\toolname}{OpenSBT }
\usetikzlibrary{shapes,arrows}

\newcommand*\circled[1]{\tikz[baseline=(char.base)]{
            \node[shape=circle,draw,inner sep=1pt] (char) {#1};}}

\definecolor{codegreen}{rgb}{0,0.6,0}
\definecolor{codegray}{rgb}{0.5,0.5,0.5}
\definecolor{codepurple}{rgb}{0.58,0,0.82}
\definecolor{backcolour}{rgb}{0.95,0.95,0.92}

\lstdefinestyle{mystyle}{
    backgroundcolor=\color{backcolour},   
    commentstyle=\color{codegreen},
    keywordstyle=\color{magenta},
    numberstyle=\tiny\color{codegray},
    stringstyle=\color{codepurple},
    basicstyle=\ttfamily\footnotesize,
    breakatwhitespace=false,         
    breaklines=true,                 
    captionpos=b,                    
    keepspaces=false,                 
    numbers=left,                    
    numbersep=3pt,                  
    showspaces=false,                
    showstringspaces=false,
    showtabs=false,                  
    tabsize=2,
    language=python,
    numbers=left,
    stepnumber=1,
}
 \setcopyright{none} 

\begin{document}

\title{OpenSBT: A Modular Framework for Search-based Testing of Automated Driving Systems}

% \author{\IEEEauthorblockN{Lev Sorokin, Tiziano Munaro, Damir Safin}
% \IEEEauthorblockA{
% \textit{fortiss, Research Institute of the Free State of Bavaria}\\
% Munich, Germany \\
% \{sorokin, munaro, safin\}@fortiss.org}
% \and
% \IEEEauthorblockN{Brian Hsuan-Cheng Liao, Adam Molin}
% \IEEEauthorblockA{\textit{DENSO AUTOMOTIVE Deutschland GmbH}\\
% Eching, Germany\\
% \{h.liao, a.molin\}@eu.denso.com}
% \\[-20.0ex]
% }
%%%%%% Centered solution
% \author{
%     \IEEEauthorblockN{Author1\IEEEauthorrefmark{1}, Author2\IEEEauthorrefmark{2}, Author3\IEEEauthorrefmark{2}, Author4\IEEEauthorrefmark{1}}
%     \IEEEauthorblockA{\IEEEauthorrefmark{1}Institution1
%     \\\{1, 4\}@abc.com}
%     \IEEEauthorblockA{\IEEEauthorrefmark{2}Institution2
%     \\\{2, 3\}@def.com}
% }
%%% For ACM format
\author{Lev Sorokin, Tiziano Munaro, Damir Safin}
\affiliation{%
  \institution{fortiss, Research Institute of the Free State of Bavaria}
  \streetaddress{Guerickestraße 25, 80805 Munich, Germany}
  \city{Munich}
  \country{Germany}}
\email{sorokin@fortiss.org, munaro@fortiss.org, safin@fortiss.org}

\author{Brian Hsuan-Cheng Liao, Adam Molin}
\affiliation{%
  \institution{DENSO AUTOMOTIVE Deutschland GmbH}
  \city{Eching}
  \country{Germany}}
\email{h.liao@eu.denso.com, a.molin@eu.denso.com}

% \author{Damir Safin}
% \affiliation{%
%   \institution{fortiss, Research Institute of the Free State of Bavaria}
%   \streetaddress{Guerickestraße 25, 80805 Munich, Germany}
%   \city{Munich}
%   \country{Germany}}
% \email{safin@fortiss.org}
\begin{abstract}
% Search-based software testing (SBT) is an effective approach for system-level testing when default testing is not possible. SBST makes use of stochastic optimization methods to identify faults in System-under Test (SUT) when the search space is high-dimensional and infinite.
% % This is due to the huge amount of testing candidates in a complex and infinite search space of the 
% % the verification of the system is not possible. 
% Especially in the automotive domain SBST is widely applied for testing Automated and Autonomous Driving Systems (AD/ADS) since it is more effective, less dangerous and time consuming than doing on-road testing.
% A lot of research is done where SBST techniques are applied and tested on AD/ADAS systems. In general, the testing environment has to be implemented from scratch, since the simulator, fitness functions and use case/scenario are X specific.
% New testing approaches are often implemented use case specific, why it requires much effort to apply testing of another SUT or with a different testing scenario. %Or they are not shared due to IP concerns and need to be reimplemented. 
Search-based software testing (SBST) is an effective and efficient approach for testing automated driving systems (ADS). However, testing pipelines for ADS testing are particularly challenging as they involve integrating complex driving simulation platforms and establishing communication protocols and APIs with the desired search algorithm.
% Nevertheless, it is challenging to set up a testing pipeline and integrate, e.g., a specific simulation environment or desired search algorithm. 
This complexity prevents a wide adoption of SBST and thorough empirical comparative experiments with different simulators and search approaches.
% In particular, it becomes complex when one does not only rely on one search algorithm or simulator for testing.
We present OpenSBT, an open-source, modular and extensible framework to facilitate the %the development, i.e., the implementation, evaluation as well as the%
SBT of ADS.
With OpenSBT, it is possible to integrate simulators with an embedded system under test, search algorithms and fitness functions for testing.
We describe the architecture and show the usage of our framework by applying different search algorithms for testing Automated Emergency Braking Systems in CARLA as well in the high-fidelity Prescan simulator in collaboration with our industrial partner DENSO. OpenSBT is available at https://git.fortiss.org/opensbt. A demo video is provided here: https://youtu.be/6csl\_UAOD\_4.
 
\end{abstract}
\settopmatter{printacmref=false}

\keywords{Search-based software testing, metaheuristics, scenario-based testing, autonomous driving, automated driving}

\maketitle

% \begin{IEEEkeywords}
% Search-based software testing, metaheuristics, scenario-based testing, autonomous driving, automated driving
% \end{IEEEkeywords}

% \end{IEEEkeywords}

\section{Introduction}
Search-based software testing (SBST) is a promising approach for virtual system-level testing of Automated Driving Systems (ADS) since it is more effective and less time-consuming than doing on-road testing \cite{marko2019, matinnejad2015}. 
 A considerable amount of research has been conducted, such as for studying the reproducability of testing results across different simulators \cite{Borg21CrossSimTesting}, investigating the transferability of virtual testing to physical testing \cite{StoccoGapTesting21} or the development of effective test case generation approaches \cite{Raja16NeuralNSGA2,Raja18NSGA2DT, Klück19Nsga2ADAS}. 
Applying SBST to ADS is challenging, as it requires several steps such as the integration of a simulation environment, definition of a fitness function, definition of a search approach, as well analysis/visualization of the test outcome. Hence, implementing a testing pipeline from scratch is complex and time-intensive \cite{AfsoonSimChallenges21}.
Further, resulting implementations are often not accessible because of intellectual property concerns or cannot be easily extended to be used with other SUTs, simulators, fitness functions or testing scenarios \cite{Raja18FeatureInteraction, Raja18NSGA2DT}. 
% Especially researchers do not  SBT experienced personal in small and medium size enterprises (SME). \cite{} %and testing engineers.
% Testing with various simulators is important since results do vary \cite{Borg21CrossSimTesting}
% [Contribution].

We present OpenSBT, a novel SBST framework for ADS which addresses the software engineering challenge of enabling a modular and flexible testing pipeline that can be applied in different use cases. OpenSBT offers the following functionalities:
% We present the framework \toolname that eases the application of SBT to AD/ADS systems with the following features: 
1) it allows to apply existing or user-defined search algorithms for testing ADS and defining fitness functions, 2) it provides an interface to integrate different SUTs and simulators without affecting other parts of the pipeline, and 3) it visualizes and analyses the test outcome. Furthermore, OpenSBT is open-source and available for both academic and commercial use.
% Our framework is considered not as a benchmarking platform of ADS, instead it enables to integrate SBT components to apply it for testing ADS, considering different configurations, such as simulators, search algorithms or fitness functions. 
We describe the architecture of OpenSBT and demonstrate its usage for testing different automated emergency breaking systems (AEB) on both, the open-source simulator CARLA \cite{Dosovitskiy17Carla} as well the high-fidelity simulator Prescan \cite{Prescan} with distinct search techniques specified by the user. Further, we report about the experience of the application of OpenSBT by our industrial partner DENSO and outline a comprehensive validation study as part of our future work.

% \item \textbf{OPTIONAL. it enables the analysis of the performance of applied SBT techniques using well known quality indicators as HV, IGD and SP [9] as required }
%\end{enumerate}

% Additionally, we provide a set of fitness/criticality functions that can be used for the application of the developed algorithm for a scenario for an ADS. We show how to define a new fitness/criticality function to use it for search. 
%and Prescan are integrated, 
%but the code base can be easily extended to support other simulators.  

%2) an algorithm that uses model fitting to identify critical scenarios. For both examples we use the following experimental setup: As the Simulator we use CARLA, as the functional scenario we consider an ego that is driving on a public city road, while a pedestrian is crossing the road. 

%Related work:

%https://arxiv.org/pdf/2206.07813.pdf

\label{sec:intro}

\section{Related Work}
\label{sec:related-work}

% Existing testing frameworks for ADS are available designed for specific use cases.
Considerable amount of research has been conducted to support virtual testing of ADS.
% (("S-TALIRO: A Tool for Temporal Logic Falsification for Hybrid Systems"))
S-Taliro~\cite{S-TaliroTuncali16}, a temporal logic falsification tool is used to generate critical test cases for ADS by falsifying safety requirements given as temporal logic specifications. The SUT needs to be provided as a Simulink model, which limits the applicability of this framework.
% ((Tools from the SBST Competitions to generate "critical" road topologies))
As part of the SBST Tools Competitions 2021/2022~\cite{GambiSBST22, PanichellaToolCompetition2021}, several tools have been developed for the generation of road topologies for testing lane keeping assist systems (LKAS). These frameworks are coupled to one simulator and one specific ADS type. On the contrary, OpenSBT can accommodate different simulators and ADS which include LKAS or collision avoidance systems such as AEBs. 
% The tools are coupled to one simulation environment and it is not possible to test other types of ADS for the violation of different safety requirements, such as collision with other dynamic/static objects. 
% On the contrary, OpenSBT can accommodate multiple simulators and multiple ADS which include lane keeping, AEB, etc.
% ((SafeBench))
SafeBench \cite{xuSafebench22} allows for testing an ADS on predefined driving scenarios and is coupled to CARLA. Also, it is not possible to integrate a different critical test-case generation algorithm or to use a SUT which must be implemented using Simulink. Further, the framework can only evaluate ADS based on reinforcement learning. Wang 
 et al. \cite{WangAdvSim21} propose a tool to generate safety-critical test cases by perturbing the behaviors of other actors. The framework focuses on Lidar-based ADS and uses one specific simulator.
%((https://chentaolue.github.io/pub-papers/ase22-adept.pdf))
The tool ADEPT \cite{WangADEPT22} simulates an ADS only in CARLA and generates adversarial attacks on deep neural networks to provoke critical scenarios.
% ((https://openpass.eclipse.org/architecture/#platform-concept))
The open-source testing platform OpenPASS \cite{OpenPASS} is similar to OpenSBT, but does not allow the integration of different simulators. \\
% ((https://people.eecs.berkeley.edu/~sseshia/pubdir/verifai-cav19.pdf))

% No open source tool exists that allow to integrate different search approaches, SUT, simulators, scenarios.

% Do we have anything else?

\section{Architecture}
\label{sec:architecture}

\begin{figure} [htp!]
    \centering
    \includegraphics[scale=0.41]{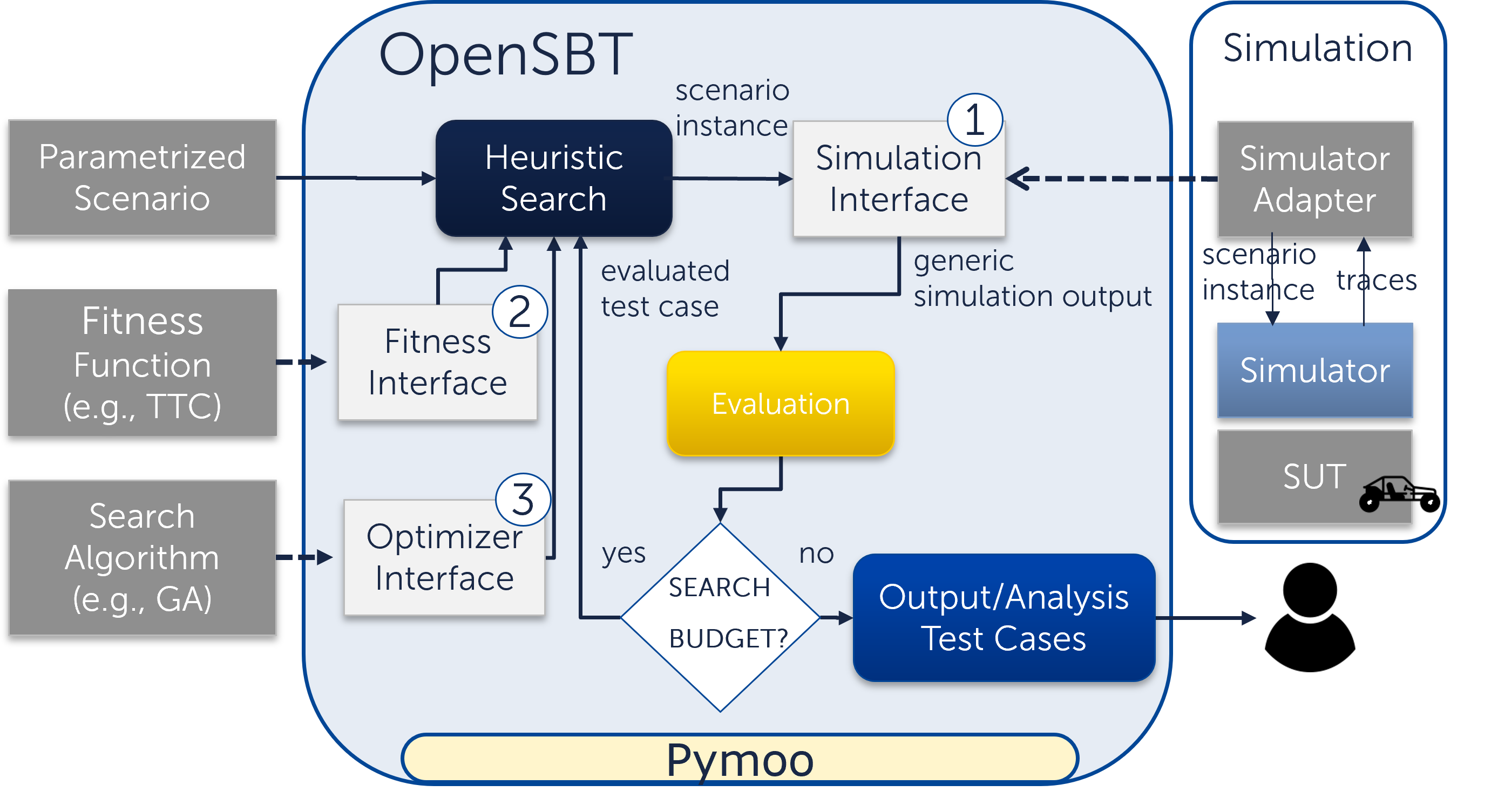}
    \caption{Architecture of \toolname}
    \label{fig:SBTg-pipeline}
\end{figure}

Our framework is based on the multi-objective optimization framework \texttt{pymoo} \cite{Blank20pymoo} and is implemented in Python. OpenSBT builds upon this framework given its modularity, extensibility and its support to numerous optimization methods. % because it has a well structured architecture and is easy to be extended to solve simulation-based optimization problems. 
In general, SBST applied to ADS requires 
1) a SUT connected to a simulation environment, 
2) a \textit{parameterized} scenario \cite{Ulbrich15scenario}, which defines the behaviour of actors/environment and provides parameters with parameter ranges to modify this behaviour,
% (a logical scenario with concrete values for each parameter is called $test\ case$)
3) a fitness function to assess the quality of a test case simulation, and 
4) a metaheuristic search algorithm.  
% The architecture of OpenSBT provides abstract definitions of the four components and interfaces that allow users to integrate their search algorithm and to define its inputs based on their needs  (s. Fig.~\ref{fig:SBTg-pipeline}).
In the following, we describe the interfaces OpenSBT provides, which allow users to integrate their search algorithm and to define its inputs based on their requirements (Fig.~\ref{fig:SBTg-pipeline}). 
% We will use the terms test case and scenario instance synonymously throughout the paper.

\textbf{Simulation/SUT.} The interface for the simulator is represented by an abstract \texttt{Simulator} class $\circled{1}$ which provides a $simulate$ method to be implemented to execute \textit{scenario instances} in a specific simulator. We denote as scenario instance a scenario which specifies concrete values for each search parameter \cite{Ulbrich15scenario}.
% The class contains parameters as \texttt{time\_step} and \texttt{simulation\_time}, which are forwarded to the concrete simulator when executing a test case. 
% The simulation of a scenario is performed by a dedicated module that connects \toolname with the Simulation.
% When $simulate$ receives a list of test cases, it passes the test cases to this module, which in turn triggers the simulator to execute the SUT on the test cases.% that connects with the simulator e.g. for the CARLA Simulator ScenarioRunner \cite{ScenarioRunner}, which is called from $simulate$ method.
Several scenario instances are passed to
 $simulate$ to allow to optimize the execution time for population-based search approaches. The result of one scenario simulation is returned in form of \texttt{SimulationOutput} instance, which holds for each actor state information for each simulation time step, as e.g., location, orientation, velocity and meta information, such as sampling rate or time step. This class can be extended with further environmental attributes, such as the position of static objects or colour of traffic lights. 

 % Consider, that since OpenSBT provides an interface for a simulator it does not impose any limitations on the simulation environment, i.e. simulators of different fidelities, whether surrogate models or physics-based simulators can be used.

\textbf{Fitness.} The fitness function assesses the quality of a test case and to should guide the search towards critical test cases. In OpenSBT, the fitness function is represented by the interface \texttt{Fitness} $\circled{2}$.  % The class a $name$, a flag whether it is minimized or maximized and an abstract 
The class provides the \texttt{eval} method for the implementation of specific evaluation instructions, which receives as input a \texttt{SimulationOutput} instance. For the optimization with multiple objectives, we use one instance of \texttt{Fitness}, to avoid redundant computation steps when optimizing different objectives.
% since the computation steps of multiple separated objective values might be redundant when representing it by several \texttt{Fitness} instances.
% Since optimal solutions output by a Pareto-based search algorithm are not necessarily failure-revealing, we have modelled a criticality function using the abstract class \texttt{Critical} in OpenSBT. 
Since optimal solutions output by a Pareto-based search algorithm are not necessarily failure-revealing, OpenSBT in addition offers the abstract class \texttt{Critical}. It allows to filter all critical test cases from the final solution set and can be used to guide the search \cite{Raja18NSGA2DT}.

% Its definition is similar to the definition of a fitness function and allows the user to express when a test case execution is critical.
% The criticality function can be used to filter all critical solutions from the final solution set, as well for guiding the search \cite{Raja18NSGA2DT}.
% To indicate when a simulation run is fault revealing or not, a criticality function can be defined in a similar way by implementing the abstract class \texttt{Critical}.
% A criticality function receives a \texttt{SimulationOutput} and returns a boolean value.
The fitness and criticality functions are domain and scenario specific, therefore they have to be defined manually by the user. Several \textit{guidelines} exist on how to define fitness and criticality functions \cite{Hauer19FitnessFF,Kolb21FitnessJunction, Shalev17RSS}. In OpenSBT are already well-applied fitness functions such as Time-to-Collision provided.

% Nevertheless, we have provided in OpenSBT fitness and criticality functions as TTC that has been often applied in SBST approaches.

% Consider that a scenario having the best fitness value is not necessarily "fault revealing", when the SUT is not faulty. For instance, if the distance to other vehicles needs to be minimized, a criticality function can declare that the distance in a simulation needs to be below a threshold the scenario to be critical.

\textbf{Algorithm.} The search algorithm is represented by the abstract class \texttt{Optimizer} \circled{3}, which provides the methods
% provides the abstract function
\texttt{init} and \texttt{run}. %which need 
% has to be implemented to use a specific search algorithm.  
 OpenSBT provides three options to specify the search algorithm: a) An existing optimization algorithm in pymoo is used and instantiated in $\texttt{init}$, b) A new algorithm is implemented by subclassing \texttt{Algorithm} in pymoo and by instantiating it in \texttt{init},
c) The \texttt{run} method of \texttt{Optimizer} is overridden and the search algorithm is implemented in run.
% The \texttt{run} method returns an extended class of type \texttt{Result} from the pymoo framework that is used for visualization/analysis of the results and the performance of the search algorithm.

\textbf{Scenario/Problem.} The scenario is specified as part of the class \texttt{ADASProblem}, which is a subclass of pymoo's class \texttt{Problem}. \texttt{ADASProblem} holds all information required to solve the underlying optimization problem such as the path to the scenario, the search space defined by the upper/lower bounds of the search variables, and instantiations of \texttt{Fitness}, \texttt{Critical}, and \texttt{Simulator}.
% is represented by an instance of \texttt{ADASProblem} \circled{4} which is a subclass of the class \texttt{Problem} in pymoo. \texttt{ADASProblem} 

% Our framework supports scenarios represented in \mbox{OpenSCENARIO} v1.2.0 (OSC), a standard introduced by ASAM\footnote{https://www.asam.net/standards/detail/openscenario} and supported by many simulators. The simulator interface can be extended to support other formats.
Our framework supports scenarios represented in \mbox{OpenSCENARIO} v1.2.0 (OSC), a standard introduced by ASAM\footnote{https://www.asam.net/standards/detail/openscenario}, as long as the simulator interface is implemented accordingly. % and supported by many simulators. The simulator interface can be extended to support other formats.
Note, that the simulator interface is not scenario-type specific.

Scenarios can be derived manually by domain experts using so-called \textit{mental models} or automatically, e.g., through clustering real driving data \cite{HauerClustering20,TUEV2022}. 
It is out of scope of this framework to provide a complete list of scenarios for testing an ADS. 
% Instead, \toolname defines an interface that allows the user to specify the scenario for testing. 
The search variables and the bounds for the variables are crucial for the successful search and can be chosen by using expert knowledge.

 % \texttt{ADASProblem} overrides internally the inherited \texttt{\_eval}. 
 % When the eval function is called during a optimization step by pymoo, first the simulate function is called to simulate test cases in the simulator and then simulation outputs are evaluated with the fitness function provided. % Further attributes that are passed are required for visualization of the search results. 
 % Further, \toolname provides the class \texttt{Experiment} to represent a testing experiment which requires as input an \texttt{ADASProblem}, an instance of \texttt{Optimizer} and \texttt{SearchConfiguration}, that holds parameters as e.g., the population size and the number of iterations.
 % The class $Experiment$ is used to maintain distinct testing experiments.
% which is defined by 
% the logical scenario represented by: 1. the names of the search variables to be modified during search,  
% 2. the search space represented by the min and max bounds of the search variables, 
% 3. the path to the functional scenario file. 
% and 4. the fitness function to assign a quality of a simulated scenario.
% The user needs to select a search algorithm and declare parameters for the search as e.g. population size, number of iterations (for population based algorithm) defined by an instance of \texttt{SearchConfiguration}. Every algorithm provides an implementation of a \texttt{DefaultSearchConfiguration} since the search configuration is algorithm specific.

\textbf{Analysis/Output.} 
After the search execution has been finished, \toolname outputs all critical and non-critical test cases and the corresponding fitness values in form of a \texttt{CSV} file and two-dimensional search space plots, i.e. \textit{design space plots}. Note, that for each pair of search variables one design space plot is generated.

Additionally, conditions over the search variables are derived using the classification and regression trees algorithm \cite{BreimanClassificationAR1984} which help to characterize the critical scenarios (e.g., ego velocity $\geq$ 20m/s\ $\wedge$\ pedestrian\ velocity\ $\geq$ 5 m/s). This information can be used to specify the operational design domain (SAE J3016) or to debug the SUT \cite{Raja18NSGA2DT}. To allow further debugging and test inspection and demonstrate to the user what has actually happened in a critical scenario, the trajectories of all scenario actors of an executed test case are visualized in an animated 2D plot. 

\section{Usage}
\label{sec:usage}

 In this section we describe the usage of OpenSBT on two different usage scenarios. A detailed tutorial of how to use OpenSBT can be found here\footnote{https://git.fortiss.org/opensbt/opensbt-core/-/tree/main/doc/jupyter}.

In the first usage scenario (S1) we use CARLA and apply OpenSBT for testing an ADS with an AEB developed in the fortiss Mobility Lab.\footnote{https://www.fortiss.org/en/research/fortiss-labs/detail/mobility-lab} We use NSGAII \cite{Deb02NSGA2} for generating the test inputs.

In the second usage scenario (S2) we use Prescan and test a Simulink-based AEB developed as part of the Automated Valet Parking use case of the FOCETA project~\cite{FOCETA} led by DENSO.  We use a different surrogate-assisted search technique NSGAII-DT~\cite{Raja18NSGA2DT}.
As testing scenario we consider an ego vehicle which drives on a straight road while an occluded pedestrian starts crossing its driving trajectory. We want to identify test cases where the AEB fails to avoid a collision with the pedestrian. Of course, more complex scenarios can be accommodated in our framework, with no impact on the usability of OpenSBT. 

In the following, we explain how to configure the search to test the SUT in both usage scenarios. 

\textbf{Simulator Integration} To integrate the SUT and the specific simulator for S1, we create the class $CarlaSimulator$ which implements the $simulate$ method of the abstract class $Simulator$ to simulate test cases.
% and return simulation outputs. 
% While the execution logic for CARLA could have been implemented directly in the $simulate$ function, it has been 
We package the execution logic for CARLA in a dedicated module\footnote{https://git.fortiss.org/opensbt/carla\_runner} instead implementing the logic in $simulate$ to avoid simulator-specific dependencies. %'s core and showcase the framework's modularity. 
By deploying multiple CARLA servers, the CARLA interface is capable of running several scenario instances in parallel. Here, scenarios which need to be evaluated are stored in a thread-safe queue. This queue is processed by a pool of worker threads, each managing one of the available CARLA servers through their TCP/IP interfaces. Once all scenarios have been evaluated, a set containing the respective \texttt{SimulationOutput}s is returned.

Besides executing embedded CARLA agents (based on the \\\texttt{AutonomousAgent}~class), the CARLA Runner module provides generic and well-used ROS and FMI interfaces\footnote{https://fmi-standard.org/tools/}.
% supported by most industry-leading tools 
Decoupling the SUT using ROS and FMI from the simulation environment further reduces the effort required to leverage SBST for systems which are provided with a ROS bridge such as Baidu's Apollo.\footnote{https://github.com/ApolloAuto/apollo/}. This example can be executed in the jupyter notebook here\footnote{https://git.fortiss.org/opensbt/opensbt-core/-/blob/update-readme/doc/jupyter/06\_Example\_CARLA.ipynb}.

For S2, we integrate Prescan by using the simulator class \textit{Prescan Simulator} and the dedicated component \texttt{Prescan\ Runner}.\footnote{https://git.fortiss.org/opensbt/prescan\_runner} The \texttt{PrescanRunner} connects to MATLAB and triggers the execution of the Simulink-based SUT in Prescan.
% In our example the SUT is provided in the same package as the module that executes the scenario in CARLA.

% In the following we describe how our tool can be used for test case generation by the following steps:
% 1) Integration of the simulator and the SUT, 2) Definition of a fitness/criticality function, 3) Definition of the search algorithm and 4) Definition of the testing scenario.

% TODO create bash script (install.sh) that does the steps above \\

% TODO linux vs. windows

% - Inherit from simulation class
% - Extension of the simulation output class
% - Override simulate\_batch function

\textbf{Fitness/Criticality Function}. 
% SBST is to be tailored to exercise the AEB and identify scenarios when it is not able to avoid a collision with a pedestrian.
We select two fitness function we will use for both usage examples: 1) \texttt{F1} outputs the minimal distance between the ego vehicle and the pedestrian, 2) \texttt{F2} calculates the velocity of the ego vehicle at the time when the minimal distance is reached. We have to subclass \texttt{Fitness} and implement its \texttt{eval} function for computing both fitness values. 
% Fig. \ref{fig:fitness} shows the implementation of the \texttt{eval} function of class \texttt{Fitness} for computing both fitness values. 
Additionally, we specify that we want to minimize \texttt{F1} and maximize \texttt{F2} to guide the search towards critical test cases.
% (Fig. \ref{fig:fitness}, line 6).
% We specify the name for each fitness function to be used later for the visualization/output of the test case generation results (Fig.. \ref{fig:fitness}, line 4). 
The criticality function is defined in a similar way as the fitness function, the \texttt{eval} method of the corresponding class returns true, when $\texttt{F1} = 0$ and $\texttt{F2} > 0$.
% fitness function has to be defined to assess the quality of single simulations.
% a two-dimensional, and as the second value the velocity at the time of the minimal distance.

% To assess the quality of a generated test case/senario we subsequently define
% % fitness function has to be defined to assess the quality of single simulations.
% a two-dimensional fitness function that outputs as the first value the minimal distance to the \textit{adversary} actor during a scenario, and as the second value the velocity at the time of the minimal distance.
% The implementation is done by implementing the $eval$ function of the interface class \texttt{Fitness} (s. Fig. \ref{fig:fitness}). We specify additionally the name for each dimension to be used in the result output.
 \vspace{-1.2pt}

% \begin{figure}[htp!]
% \begin{lstlisting}[python]
% class FitnessMinDistanceVelocity(Fitness):
%  @property
%  def min_or_max(self):
%     return "min", "max"
    
%  def eval(self, simout: SimulationOutput):
%     tr_ego = simout.location["ego"]
%     tr_pred = simout.location["adversary"]
%     i_min = np.argmin(geo.distPair(tr_ego, tr_pred))
%     distance = geo.distPair(tr_ego, tr_pred)[i_min]
%     speed = simout.speed["ego"][i_min]
%     return (distance, speed)
% \end{lstlisting}
% \caption{Fitness function implementation}
% \label{fig:fitness}
% \end{figure}

% In our framework we have implemented the aforementioned fitness function as well a function that is using the well adopted safety measure Time-To Collision \cite{Horst94TTC}, to detect near-to-collision situations.

\textbf{Search Algorithm}. NSGA-II used in S1 is already implemented in pymoo, therefore we only need to implement the $init$ method of \texttt{Optimizer} using the method a) as explained in Section \ref{sec:architecture} - \textit{algorithm}.
 For S2, we implement NSGAII-DT in the optimizer class using the method c), as the search algorithm is not implemented in pymoo. 
% To use a new algorithm, we could create a class that inherits from the class \texttt{Algorithm} in pymoo instead and assign its instance to the \texttt{Optimizer}'s algorithm attribute.

\textbf{Problem}. In both usage scenarios three input variables are involved in the search process: \texttt{EgoSpeed}: the velocity of ego, \texttt{PedSpeed}: the velocity of the pedestrian, \texttt{PedDist}: the distance to the ego vehicle, when the pedestrian starts walking. We set the parameter bounds in such a way that collisions can occur.
% \texttt{EgoSpeed} and \texttt{PedSpeed} based on commonly used values and choose for the upper bound for \texttt{PedDist} the initial distance.
For S1, we provide an OSC file where the behaviour of the pedestrian/environment is specified and the described variables defined. 
For S2, we pass a different Prescan-specific file in the \texttt{pb} format, as Prescan has a limited support for OSC.

For both usage scenarios, we pass the search variables, the search bounds, and the scenario file to \texttt{ADASProblem} as done for S1 in Fig. \ref{fig:problem}. We configure the simulator (line 10) and pass the fitness/criticality function we defined before (line 9,11).

\begin{figure} [!htp!]
\begin{lstlisting}[language=python]
from simulator.carla_simulator import CarlaSimulator

problem = ADASProblem(
    scenario_path = "/tmp/PedestrianCrossing.xosc",
    simulation_variables = ["PedSpeed",
       "EgoSpeed","PedDist"],
    xl = [0.5, 1,0], # [m/s] [m/s] [m]
    xu = [3, 22,60], 
    fitness_fnc = FitnessMinDistanceVelocity(),
    simulate_fnc = CarlaSimulator.simulate,
    critical_fnc = CriticalAdasDistanceVelocity(),
    problem_name = "PedestrianCrossingCarla")

exp = Experiment(name="1",
         problem=problem,   
         algorithm=AlgorithmType.NSGAII,
         search_config=DefaultSearchConfiguration())
\end{lstlisting}
\vspace*{-3mm}
\caption{Problem definition for the first usage scenario}
\label{fig:problem}
\end{figure}
% We pass simulation config parameter together with the simulation function.
% In the next steps we need to define the algorithm to be used for  
% the \texttt{Experiment} 
% - Define ADAS problem
% - Define search algorithm
% - Define search configuration
We configure the search algorithm and set the search configuration (e.g. search time, population size) by instantiating the \texttt{Experiment} class (for S1 s. Fig.~\ref{fig:problem}, line 14-17).
The experiment definition for S2 is similar, the only difference is that NSGAII-DT is specified instead NSGA-II.

% The datastructure 
% \texttt{DefaultSearchConfiguration} can be extended to hold other algorithm specific search parameters. 
% \begin{figure}[h!]
% \begin{lstlisting}[language=python]

% \end{lstlisting}
% \caption{Experiment definition}
% \label{fig:experiment}
% \end{figure}
\textbf{Search Execution}. The search can be executed from a python script or via the console. To use OpenSBT via console, for S1, we start the search by executing the command \texttt{run.py\ -e\ 1} in the command line,
% In the current implementation of \toolname the user has to define experiments via code. 
% Further, it is possible to adjust defined experiments using the command line.
% e.g., the search algorithm, the search space or a search configuration.
where the flag \texttt{e} holds the experiment name. Further we can modify experiments. To restrict the search time to two hours and set the population size to 50 we need to execute \texttt{python run.py -e 1 -n 50 -t "02:00:00"}. A description of further flags and how to run OpenSBT directly from a python script can be found in OpenSBT's documentation.

\begin{figure} [h]
    \centering
    \includegraphics[scale=0.25]{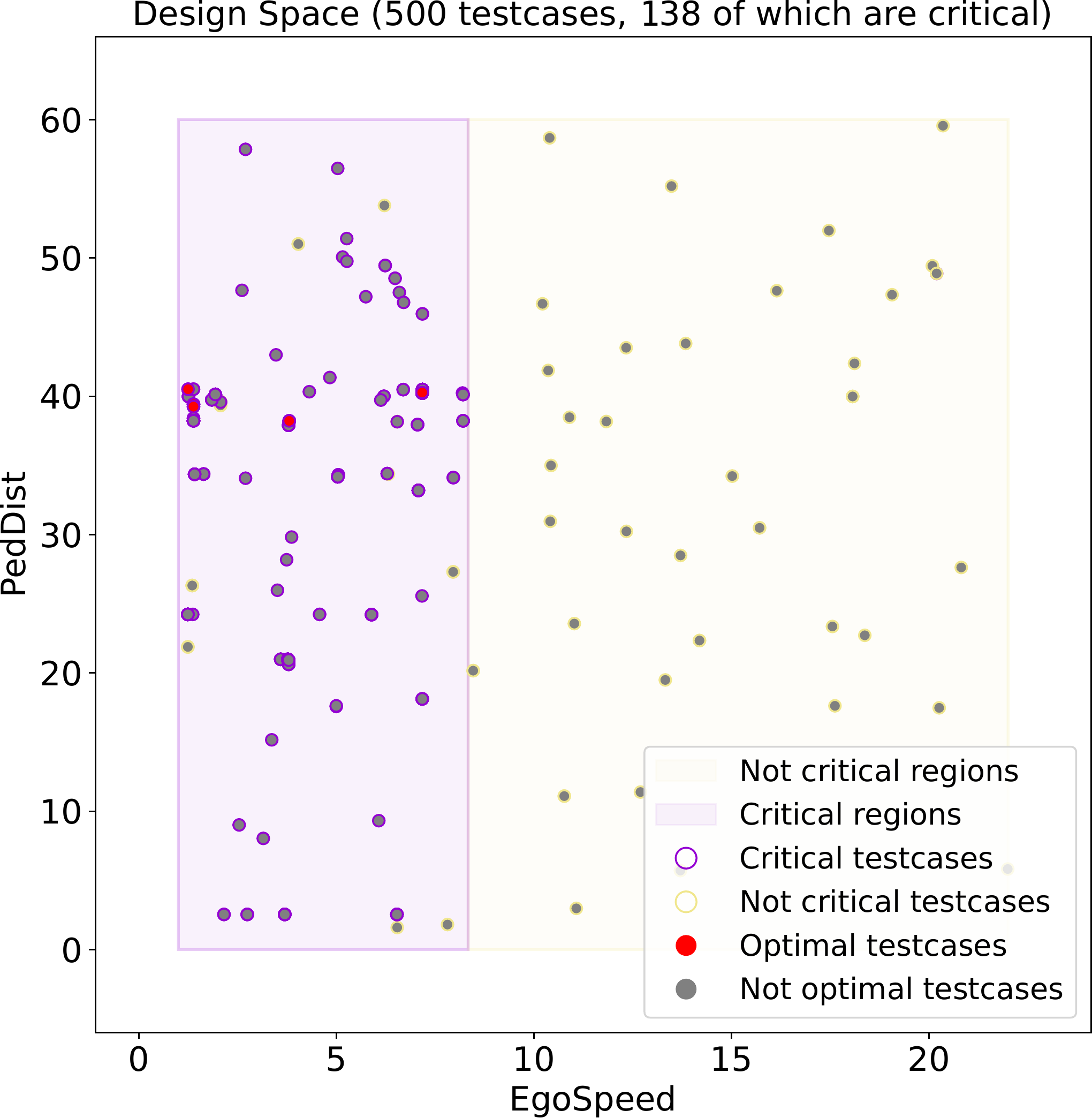}
    \caption{Test input space plot with derived criticality condition $EgoSpeed < 8.33 m/s$ (purple box) by decision tree learning}% Region 1: $0 \leq v_e \leq 1 \wedge 0 \leq v_e \leq 1.3$, Region 2: $  2 \leq v_e \leq 2.5\wedge 1.7 \leq a \leq 2.2$, Region 3: $  3.5 \leq v_e \leq 5 \wedge 0 \leq v_p \leq 3$ %
    \label{fig:design-space}
\end{figure}

% We can modify the search by using the following parameters: \texttt{n}: for setting the population-size, \texttt{t}: the search time, \texttt{a}: the search algorithm.
% (s. Fig. \ref{fig:cli}).
% the search with 1) a specific search algorithm, 2) a specific search configuration and 3) a specific search interval. 
% For instance, the following command  runs search for the defined experiment in \ref{fig:experiment} know as experiment 1 with NSGAII (= algorithm 1). The tested scenario is a pedestrian that crosses the lane of the ego car. The search parameters that are varied are the speed of the ego vehicle and the speed of the pedestrian. The search time is set to one hour.
%  \vspace{-1.2pt}
% \begin{figure}
%     \centering
% \begin{lstlisting}[language=shell]
% python run.py -e 1 -a 1 -min 0 0 -max 30 2 -m "EgoSpeed" "PedSpeed" -t "01:00:00"
% \end{lstlisting}
% \vspace*{-3mm}
%     \caption{Search execution}
%     \label{fig:cli}
% \end{figure}
% Passing the \texttt{-h} flag returns a list of all available options. 
When the search has terminated, results are stored in the \texttt{results} directory. The generated visualizations can be used to determine the domain of the search space in which the SUT behaves faulty. For instance, in Fig.~\ref{fig:design-space} is the design space plot from S1 for the variables \texttt{PedSpeed} and \texttt{EgoSpeed} depicted, which contains one derived condition when the SUT behaves critical.
\section{Evaluation}
% Usage section -> already preliminary evaluation "

In the previous section we have presented two usage examples which serve as a preliminary evaluation of OpenSBT.
In addition, we have provided OpenSBT to our industrial partner DENSO.
% As a preliminary evaluation experiment we have provided OpenSBT to our industrial partner DENSO.
DENSO engineers have applied OpenSBT to benchmark different evolutionary algorithms such as (Multi-Objective) Particle Swarm Optimization on an AEB system. The integration of these algorithms has been done similarly to that of NSGAII-DT. Based on the feedback by DENSO engineers, testing the AEB with different search techniques and comparing the test outcomes have been facilitated by OpenSBT. Given the positive feedback, we are planning a more comprehensive user study by means of replication experiment \cite{BasiliExperimental93} with computer science master students in the scope of a practical course. Here, the idea is to have different groups of 4-5 participants, where the groups will perform different SBST tasks, such as testing an AEB or an LKAS. One part of the participants of a group will perform  testing by engineering a testing pipeline without the support of OpenSBT. Another part will use OpenSBT for testing.  By interviewing study candidates, we will try to make sure that selected participants have similar software engineering skills.
To evaluate OpenSBT w.r.t. its usability/flexibility we are planning to use the following metrics for each of the groups: The time required a) to perform testing the AEB with a predefined test configuration, b) to modify/extend the testing pipeline when using a different test configuration with/out using OpenSBT. The times will be collected based on time-tracking version control issues. To mitigate incorrect time tracking, we will perform in addition a qualitative evaluation and query participants about the perceived complexity implementing the testing pipeline.

\section{Conclusion and future work}
\label{sec:conclusion}

We have presented a modular framework which tackles the engineering challenge of setting up a testing pipeline that is compatible with different simulators, search algorithms and fitness functions.
% It provides generic interfaces for the integration of simulators, search algorithms, and fitness functions, as well (driving) scenarios.
% facilitating the application of SBT in different contexts.
% and applied on individual SUTs and testing scenarios.
% The framework can be extended by researchers implementing new search algorithms, fitness functions.
% . e.g. the simulation output can be extended to hold roll and pitch angles.
% ...
% Limitations...
We have conducted a preliminary evaluation of OpenSBT on two usage scenarios with different search approaches and simulators.
A comprehensive validation study by means of a replication project is part of our future work.
% We have described the usage of OpenSBT on two usage scenarios with different search approaches and simulators.
% While it still requires some effort to integrate a different simulator, the provided interface allows to specify concrete simulator adapters to connect OpenSBT to an arbitrary simulator.
% We have discussed the interfaces OpenSBT provides, e.g., the interface that allows using OpenSBT with simulators different from the ones currently supported (CARLA and PreScan).
% We have provided two simulation adapters to integrate the CARLA and Prescan simulator into OpenSBT, still some effort is required to integrate a different simulator for the execution of scenarios.
% OpenSBT users are thus invited to contribute to the collection of implementations of such simulator \textit{adapters}.
Also, we are working on easing the integration of existing search algorithms implemented in different optimization frameworks than pymoo. Also, we consider to extend OpenSBT by a graphical user interface (GUI) to create and modify experiments and to display the results of the search in the GUI. We believe that OpenSBT will contribute both to the development as well application of effective virtual testing approaches to increase the safety of ADS.

% TODO evt. RSS erwähnen

% \textbf{TODO mention: SUT is still highly coupled to the simulator, for Prescan we assume to have a simulink node. For CARLA we assume it is provided as a ROS node.}

% TODOs
% TODO integrating of different algorithm by (class) name

\section*{Acknowledgments}
\includegraphics[scale=0.018]{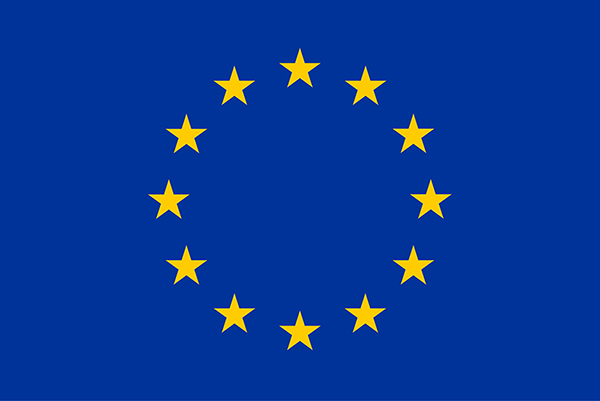} 
This paper has received funding from the European Union’s Horizon 2020 research and innovation programme under grant agreement No 956123.

\printbibliography

@inproceedings{Kolb21FitnessJunction,
  author={Kolb, Nicola and Hauer, Florian and Pretschner, Alexander},
  booktitle={
  IEEE International Intelligent Transportation Systems Conference (ITSC)}, 
  title={Fitness Function Templates for Testing Automated and Autonomous Driving Systems in Intersection Scenarios}, 
  year={2021},
  volume={},
  number={},
  pages={217-222}
  }

@inproceedings{Hauer19FitnessFF,
  title={Fitness Functions for Testing Automated and Autonomous Driving Systems},
  author={Florian Hauer and Alexander Pretschner and Bernd Holzm{\"u}ller},
  booktitle={International Conference on Computer Safety, Reliability, and Security},
  year={2019}
}

@inproceedings{Raja18NSGA2DT,
  author={Ben Abdessalem, Raja and Nejati, Shiva and C. Briand, Lionel and Stifter, Thomas},
  booktitle={40th International Conference on Software Engineering (ICSE)}, 
  title={Testing Vision-Based Control Systems Using Learnable Evolutionary Algorithms}, 
  year={2018},
  volume={},
  number={},
  pages={1016-1026}}

@inproceedings{Raja16NeuralNSGA2,
author = {Ben Abdessalem, Raja and Nejati, Shiva and Briand, Lionel C. and Stifter, Thomas},
title = {Testing Advanced Driver Assistance Systems Using Multi-Objective Search and Neural Networks},
year = {2016},
booktitle = {Proceedings of the 31st IEEE/ACM International Conference on Automated Software Engineering},
pages = {63–74},
numpages = {12},
series = {ASE 2016}
}

@ARTICLE{Blank20pymoo,
    author={J. {Blank} and K. {Deb}},
    journal={IEEE Access},
    title={pymoo: Multi-Objective Optimization in Python},
    year={2020},
    volume={8},
    number={},
    pages={89497-89509},
}

@inproceedings{Dosovitskiy17Carla,
  title = {{CARLA}: {An} Open Urban Driving Simulator},
  author = {Alexey Dosovitskiy and German Ros and Felipe Codevilla and Antonio Lopez and Vladlen Koltun},
  booktitle = {Proceedings of the 1st Annual Conference on Robot Learning},
  pages = {1--16},
  year = {2017}
}

@INPROCEEDINGS{Klück19Nsga2ADAS,
  author={Klück, Florian and Zimmermann, Martin and Wotawa, Franz and Nica, Mihai},
  booktitle={19th International Conference on Software Quality, Reliability and Security}, 
  title={Genetic Algorithm-Based Test Parameter Optimization for ADAS System Testing}, 
  year={2019},
  volume={},
  number={},
  pages={418-425}
}

@INPROCEEDINGS {Borg21CrossSimTesting,
author = {M. Borg and R. Abdessalem and S. Nejati and F. Jegeden and D. Shin},
booktitle = {14th IEEE Conference on Software Testing, Verification and Validation (ICST)},
title = {Digital Twins Are Not Monozygotic – Cross-Replicating ADAS Testing in Two Industry-Grade Automotive Simulators},
year = {2021},
volume = {},
pages = {383-393}
}

@misc{Shalev17RSS,
author = {Shalev-Shwartz, Shai and Shammah, Shaked and Shashua, Amnon},
  
  title = {On a Formal Model of Safe and Scalable Self-driving Cars},
  
  publisher = {arXiv},
  
  year = {2017},
  
  copyright = {arXiv.org perpetual, non-exclusive license}
}

@conference{marko2019,
    author={Nadja Marko. and Jonas Ruebsam. and Andreas Biehn. and Hannes Schneider.},
    title={Scenario-based Testing of ADAS - Integration of the Open Simulation Interface into Co-simulation for Function Validation},
    booktitle={9th International Conference on Simulation and Modeling Methodologies, Technologies and Applications},
    year={2019},
    pages={255-262}
}

@article{matinnejad2015,
    title = {Search-based automated testing of continuous controllers: Framework, tool support, and case studies},
    journal = {Information and Software Technology},
    volume = {57},
    pages = {705-722},
    year = {2015},
    author = {Reza Matinnejad and Shiva Nejati and Lionel Briand and Thomas Bruckmann and Claude Poull},
}

@ARTICLE{Deb02NSGA2,
  author={Deb, K. and Pratap, A. and Agarwal, S. and Meyarivan, T.},
  journal={IEEE Transactions on Evolutionary Computation}, 
  title={A fast and elitist multiobjective genetic algorithm: NSGA-II}, 
  year={2002},
  volume={6},
  number={2},
  pages={182-197}}

@online{OpenSBT,
  doi = {},
  url = {https://git.fortiss.org/opensbt/opensbt-core
  },
  year = {2023-01-08},
}

@INPROCEEDINGS{Raja18FeatureInteraction,
  author={Ben Abdessalem, Raja and Panichella, Annibale and Nejati, Shiva and Briand, Lionel C. and Stifter, Thomas},
  booktitle={33rd IEEE/ACM International Conference on Automated Software Engineering (ASE)}, 
  title={Testing Autonomous Cars for Feature Interaction Failures using Many-Objective Search}, 
  year={2018},
  volume={},
  number={},
  pages={143-154},
  }

@INPROCEEDINGS{Ulbrich15scenario,
  author={Ulbrich, Simon and Menzel, Till and Reschka, Andreas and Schuldt, Fabian and Maurer, Markus},
  booktitle={18th International Conference on Intelligent Transportation Systems}, 
  title={Defining and Substantiating the Terms Scene, Situation, and Scenario for Automated Driving}, 
  year={2015},
  volume={},
  number={},
  pages={982-988}}

@INPROCEEDINGS{HauerClustering20,
  author={Hauer, Florian and Gerostathopoulos, Ilias and Schmidt, Tabea and Pretschner, Alexander},
  booktitle={2020 IEEE Intelligent Vehicles Symposium (IV)}, 
  title={Clustering Traffic Scenarios Using Mental Models as Little as Possible}, 
  year={2020},
  volume={},
  number={},
  pages={1007-1012}}

@article{StoccoGapTesting21,
  title={Mind the Gap! A Study on the Transferability of Virtual Versus Physical-World Testing of Autonomous Driving Systems},
  author={Stocco, Andrea and Pulfer, Brian and Tonella, Paolo},
  journal={IEEE Transactions on Software Engineering},
  year={2021},
  volume={49},
  pages={1928-1940}
}

@online{Prescan,
  title={Prescan},
  doi = {},
  url = {
    https://plm.sw.siemens.com/de-DE/simcenter/autonomous-vehicle-solutions/prescan/
  },
  year = {2023-04-24},
}

@misc{xuSafebench22,
      title={SafeBench: A Benchmarking Platform for Safety Evaluation of Autonomous Vehicles}, 
      author={Chejian Xu and Wenhao Ding and Weijie Lyu and Zuxin Liu and Shuai Wang and Yihan He and Hanjiang Hu and Ding Zhao and Bo Li},
      year={2022},
      archivePrefix={arXiv},
      primaryClass={cs.RO}
}

@INPROCEEDINGS {PanichellaToolCompetition2021,
author = {S. Panichella and A. Gambi and F. Zampetti and V. Riccio},
booktitle = {14th International Workshop on Search-Based Software Testing},
title = {SBST Tool Competition 2021},
year = {2021},
volume = {},
issn = {},
pages = {20-27}
}

@online{OpenPASS,
  title={OpenPASS},
  doi = {},
  url = {
https://openpass.eclipse.org/architecture/#platform-concept
  },
  year = {2023-04-24},
}

@INPROCEEDINGS{S-TaliroTuncali16,
  author={Tuncali, Cumhur Erkan and Pavlic, Theodore P. and Fainekos, Georgios},
  booktitle={19th ITSC}, 
  title={Utilizing S-TaLiRo as an automatic test generation framework for autonomous vehicles}, 
  year={2016},
  volume={},
  number={},
  pages={1470-1475}}

@inproceedings{BreimanClassificationAR1984,
  title={Classification and Regression Trees},
  author={L. Breiman and Jerome H. Friedman and Richard A. Olshen and C. J. Stone},
  year={1984}
}

@article{WangAdvSim21,
  author       = {Jingkang Wang and
                  Ava Pun and
                  James Tu and
                  Sivabalan Manivasagam and
                  Abbas Sadat and
                  Sergio Casas and
                  Mengye Ren and
                  Raquel Urtasun},
  title        = {AdvSim: Generating Safety-Critical Scenarios for Self-Driving Vehicles},
  journal      = {CoRR},
  volume       = {abs/2101.06549},
  year         = {2021},
  bibsource    = {dblp computer science bibliography, https://dblp.org}
}

@inproceedings{GambiSBST22,
author = {Gambi, Alessio and Jahangirova, Gunel and Riccio, Vincenzo and Zampetti, Fiorella},
title = {SBST Tool Competition 2022},
year = {2023},
address = {New York, NY, USA},
booktitle = {Proceedings of the 15th Workshop on Search-Based Software Testing},
pages = {25–32},
numpages = {8},
series = {SBST '22}
}

@INPROCEEDINGS{AfsoonSimChallenges21,
  author={Afzal, Afsoon and Katz, Deborah S. and Le Goues, Claire and Timperley, Christopher S.},
  booktitle={2021 14th IEEE Conference on Software Testing, Verification and Validation (ICST)}, 
  title={Simulation for Robotics Test Automation: Developer Perspectives}, 
  year={2021},
  volume={},
  number={},
  pages={263-274}}

@report{TUEV2022,
   author = {Christoph Miethaner and Jann-Eve Stravesand},
   title = {Virtual homologation of an ALKS according to UNECE R157},
   year = {2022},
}

@inproceedings{WangADEPT22,
author = {Wang, Sen and Sheng, Zhuheng and Xu, Jingwei and Chen, Taolue and Zhu, Junjun and Zhang, Shuhui and Yao, Yuan and Ma, Xiaoxing},
title = {ADEPT: A Testing Platform for Simulated Autonomous Driving},
year = {2023},
booktitle = {37th IEEE/ACM International Conference on Automated Software Engineering},
articleno = {150},
numpages = {4},
series = {ASE '22}
}

@online{FOCETA,
  title={FOCETA},
  doi = {},
  url = {
http://www.foceta-project.eu/
  },
  year = {2023-04-24},
}

@InProceedings{BasiliExperimental93,
author="Basili, Victor R.",
title="The experimental paradigm in software engineering",
booktitle="Experimental Software Engineering Issues: Critical Assessment and Future Directions",
year="1993",
pages="1--12",
}
\end{document}